\documentclass[useAMS,usenatbib,twocolumn,preprintnumbers,nofootinbib]{revtex4}
\usepackage{epsfig}
\input psfig.sty

\usepackage{graphicx}% Include figure files
\usepackage{dcolumn}% Align table columns on decimal point
\usepackage{bm}% bold math
\usepackage{epsfig}
\usepackage{graphicx}% Include figure files
\usepackage{dcolumn}% Align table columns on decimal point
\usepackage{bm}% bold math
\usepackage[english]{babel}

\begin{document}

\title{Probing the distance-duality relation with high-$z$ data} 

\author{R. F. L. Holanda$^{1,2,3}$}\email{holanda@uepb.edu.br}

\author{V. C. Busti$^{4,5}$}\email{viniciusbusti@gmail.com}

\author{F. S. Lima$^2$}\email{limasdl@bol.com.br}

\author{J. S. Alcaniz$^6$}\email{alcaniz@on.br}

\affiliation{$^1$Departamento de F\'{\i}sica, Universidade Estadual da Para\'{\i}ba, 58429-500, Campina Grande - PB, Brasil}

\affiliation{$^2$Departamento de F\'{\i}sica, Universidade Federal de Campina Grande, 58429-900, Campina Grande - PB, Brasil}

\affiliation{$^3$Departamento de F\'{\i}sica, Universidade Federal do Rio Grande do Norte, 59072-970, Natal - RN, Brasil}

\affiliation{$^4$Departamento de F\'{\i}sica Matem\'{a}tica, Instituto de F\'{\i}sica, Universidade de S\~{a}o Paulo, 
CEP 05508-090, S\~{a}o Paulo - SP, Brasil}

\affiliation{$^5$Departament of Physics and Astronomy, University of Pennsylvania, Philadelphia, PA 19104, USA}

\affiliation{$^6$Departamento de Astronomia, Observat\'orio Nacional, 20921-400, Rio de Janeiro - RJ, Brasil}

\date{\today}

\begin{abstract}

Measurements of strong gravitational lensing jointly with type Ia supernovae (SNe Ia) observations have been used to test the validity of the cosmic distance duality relation (CDDR), $D_L(z)/[(1+z)^2D_A(z)]=\eta=1$, where $D_L(z)$ and $D_A(z)$ are the luminosity and the angular diameter distances to a given redshift $z$, respectively. However, several lensing systems lie in the interval   $1.4 \leq z \leq 3.6$ i.e., beyond the redshift range  of current SNe Ia compilations ($z \approx 1.50$),  which prevents this kind of test to be fully explored. In this paper, we circumvent this problem by testing the CDDR considering  observations of strong gravitational lensing along with SNe Ia and { a subsample from} the latest gamma-ray burst distance modulus data, whose redshift range is $0.033 \leq z \leq 9.3$. {  We parameterize their luminosity distances  with a second degree polynomial function and search for possible deviations from the CDDR validity by using four different $\eta(z)$ functions:  $\eta(z)=1+\eta_0z$,  $\eta(z)=1+\eta_0z/(1+z)$,  $\eta(z)=(1+z)^{\eta_0}$ and  $\eta(z)=1+\eta_0\ln(1+z)$. Unlike previous tests done at redshifts lower than $1.50$, the likelihood for $\eta_0$ depends strongly on the $\eta(z)$ function considered, but we find no significant deviation from the CDDR validity ($\eta_0=0$). However, our analyses also point to the fact that caution is needed when one fits data in higher redshifts to test the CDDR as well as a better understanding of the mass distribution of lenses also is required for more accurate results.}

\end{abstract}

%\pacs{98.80.-k, 98.80.Es, 98.65.Cw}

\maketitle
  
\section{Introduction}

In recent years, several authors have proposed methods to test the so-called cosmic distance duality relation (CDDR), which is derived from Etherington reciprocity theorem \cite{Eth}. In the astronomical framework, considering that photons number is conserved over the cosmic evolution and that they follow null (unique) geodesics, the CDDR is expressed as 
\begin{equation}
{D_L(z) \over  D_{A}(z) (1 + z)^{2}} =\eta=  1\;,
\end{equation}
where $D_L$ and $D_A$ are the luminosity and angular diameter distances to a given redshift $z$ respectively. The CDDR plays an essential role in observational cosmology and any departure from it could point { to unaccounted systematic errors as photon absorption by dust or even exciting new fundamental physics, such as, photon coupling with particles beyond the standard model of particle physics, variation of fundamental constants,  among others} (for a detailed discussion, see \cite{Ell} and \cite{Avg}). 

In order to test the Eq. (1) different tests involving observations of the luminosity distances of type Ia supernovae, angular diameter distances of galaxy clusters, gas mass fractions of galaxy clusters, baryon acoustic oscillations, the blackness of the cosmic microwave background and gamma-ray bursts  data have been performed. Basically, these methods may be cosmological model-dependent or independent. The former put more restrictive limits on possible deviations from Eq.(1), but they are less ideal methods since their results are valid for a specific cosmological model, generally the $\Lambda$CDM model (see \cite{Avg,Dis1}). The cosmological model-independent methods use only  astrophysical quantities and, as a weak point, the results have larger errors (see Section IV and Refs. \cite{dis2,hga,Li,gha,santos,Hol2,Nair}).  Moreover, these latter methods have been limited by the type Ia supernovae redshifts used ($z \approx 1.50$). Thus, although the results from both approaches do not show significant  deviations from CDDR validity, it is still important to propose new methods with different astronomical observations and redshift range, preferably at high redshifts, to validate the whole cosmological framework as well as to detect unexpected behavior or even systematic errors. Very recently, observations of strong gravitational lenses systems were used to perform tests of the CDDR \cite{Hol,Lia}. The crucial point here is that the sources of these systems enable to test the CDDR at higher redshifts ($z>2$). { In this paper we explore further possible constraints on the CDDR with this kind of observation.

Strong gravitational lensing is an important effect arising from Einstein's theory of general relativity, which has played a very important role as a probe of the nature of dark matter in the universe as well as in tests of cosmological models \cite{SGL}.} A known quantity is the Einstein radius, which varies with cosmological models via the ratio of the angular diameter distances  between lens/source and observer/source. Basically, it occurs when the source (s), the lens (l) and the observer (o) are well aligned with the observer-source direction \cite{Sch}. The Einstein radius $\theta_E$, under the simplest assumption describing the mass distribution in lenses, the singular isothermal sphere (SIS), is given by
\begin{equation}
 \label{thetaE_SIS}
{{\theta}_E}=4{\pi}{\frac{D_{A_{ls}}} {D_{A_s}}}{\frac{{\sigma}^2_{SIS}} {c^2}},
\end{equation}
where $c$ is the speed of light, $D_{A_{ls}}$ and $D_{A_{s}}$ are the angular diameter distances between lens-source and source-observer, respectively, and $\sigma_{SIS}$ is the velocity dispersion due to lens mass distribution. { The Einstein radius measurements jointly with SNe Ia observations and a deformed CDDR, $D_L D_A^{-1}(1+z)^{-2}=\eta(z)$, have been used recently to perform constraints on $\eta(z)$ functions (see next section for details).} For instance, by considering  $\eta(z)=1+\eta_0z/(1+z)$, the authors of  \cite{Hol}  used 95 data points from 118 strong gravitational lensing (SGL) systems obtained by \cite{Cao},  580 type Ia supernovae (SNe Ia) from Union2.1 \cite{Suz}, the latest results from {\it Planck} collaboration \cite{Ade}  and WMAP9 satellite \cite{Hin} to put limits on $\eta_0$. In \cite{Lia} was proposed a cosmological-model independent test (only a flat universe was assumed) by assuming $\eta(z)=1+\eta_0z$ and using 60 SGL-SNe Ia pairs from \cite{Bet} and \cite{Cao}. The results from both works confirmed the validity of the CDDR ($\eta_0 \approx 0$ into 1 $\sigma$ c.l.). { However, in Ref.\cite{Lia} there is a large interval allowed for $\eta_0$: $-0.1 \leq \eta_0 \leq 0.8$ at 2$\sigma$ c.l..}

On the other hand, the sample from \cite{Cao} has SGL systems that reach $z$ up to 3.6 and, consequently, a basic limitation of the analyses quoted above is the impossibility of using the entire SGL sample. The reason is very simple: to perform the test it is necessary to find SNe Ia and SGL systems at identical redshifts, but several lenses of the SGL data  lie at $z > 1.50$, which is nearly the maximum redshift of the SNe Ia in the Union2.1 compilation. Therefore, many SGL systems  were excluded in the previous analyses. On the other hand, since one of the main goal of this kind of analysis is to test the validity of the  CDDR at different regimes, reaching higher redshifts is of fundamental importance to put limits on a possible $\eta(z)$ evolution.

In this paper,  based on the method proposed by  { \cite{Lia} (see next section for details)}, we test the CDDR validity up to high redshifts by using all the 118 SGL systems from  \cite{Cao}, 580 SNe Ia from the Union2.1 compilation and { a subsample from} the latest gamma-ray burst distance module data \cite{Dem}. Gamma-ray bursts (GRBs) are the most energetic explosions in the Universe being detectable up to very high redshifts. The new data corresponds to 162 points in the redshift range $0.033 < z < 9.3$, therefore, beyond the  SNe Ia redshifts. { However, we excluded 15 GRBs that are beyond the redshift of the SGL systems ($z>3.6$). Thus, we add to the SNe Ia 580 Union2.1 data 147 gamma-ray bursts  resulting in  727 luminosity distances. Since we need to have luminosity and angular diameter distances at identical redshifts we fit the luminosity distances with a second degree polynomial fit (see Fig.1). From the theoretical viewpoint, we consider four $\eta(z)$ functions to test the Eq. (1) (see next section).}

This paper is organized as follows: Section II briefly revises the method proposed by Ref. \cite{Lia}. In Section III we present the samples used in the analysis. In Section IV we show and discuss the main results of this paper. We summarize the results of this work in Section V.

\section{Method}

The method to test the CDDR validity proposed by \cite{Lia}  uses SGL systems and SNe Ia observations and does not depend on  assumptions concerning the details of cosmological model, only a flat universe is assumed.  

Firstly, let us consider  Eq. (2) and define the observational quantity: 
\begin{equation}
\label{D}
D={\frac{D_{A_{ls}}} {D_{A_s}}}=\frac{{\theta}_E c^2}{4{\pi} \sigma^2_{SIS}}.
\end{equation}
On the other hand, in a flat universe, the comoving distance $r_{ls}$ is given by \cite{Bal}
\begin{equation}
r_{ls}=r_s-r_l. 
\end{equation}
Now, using $r_s=(1+z_s)D_{A_s}$, $r_l=(1+z_l)D_{A_l}$  and $r_{ls}=(1+z_s)D_{A_{ls}}$, we find
\begin{equation}
\label{d2}
D= 1 - \frac{(1+z_l)D_{A_{l}}}{(1+z_s)D_{A_{s}}}.
\end{equation}
Finally, using the deformed CDDR, such as, $D_L D_A^{-1}(1+z)^{-2} = \eta(z) $, the above expression can be written as
\begin{equation}
\label{d3}
\frac{(1+z_s)\eta(z_s)}{(1+z_l)\eta(z_l)}= (1-D)\frac{D_{L_s}}{D_{L_l}}.
\end{equation}
{ In \cite{Lia} the CDDR was defined as $D_L^{-1}D_A(1+z)^2=\eta(z)=1$ and the final expression  is slightly different of our Eq.(\ref{d3}).} Then,  with  SGL measurements and the luminosity distances to the lens and source of each system known, it is possible to study the evolution of the $\eta(z)$ function and test the CDDR. { In order to test the CDDR validity, we consider the following CDDR deformations:}

\begin{itemize}
\item P1: $\eta(z)=1+\eta_0 z$

\item P2: $\eta(z)=1+\eta_0 z/(1+z)$

\item P3: $\eta(z)=(1+z)^{\eta_0}$

\item P4: $\eta(z)=1+ \eta_0 \ln(1+z)$.
\end{itemize}

\begin{figure}
\centering
\includegraphics[width=0.45\textwidth]{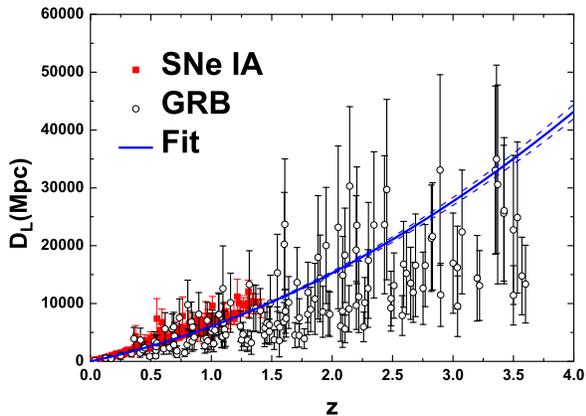}
\caption{The luminosity distances from the 580 SNe Ia (filled red circles) and 147 gamma-ray bursts (GRBs) data (open black circles). The solid and dashed blue lines are the polynomial fit and the 1$\sigma$ error from our fit, respectively.}
\end{figure}

\begin{figure*}
\centering
\includegraphics[width=0.46\textwidth]{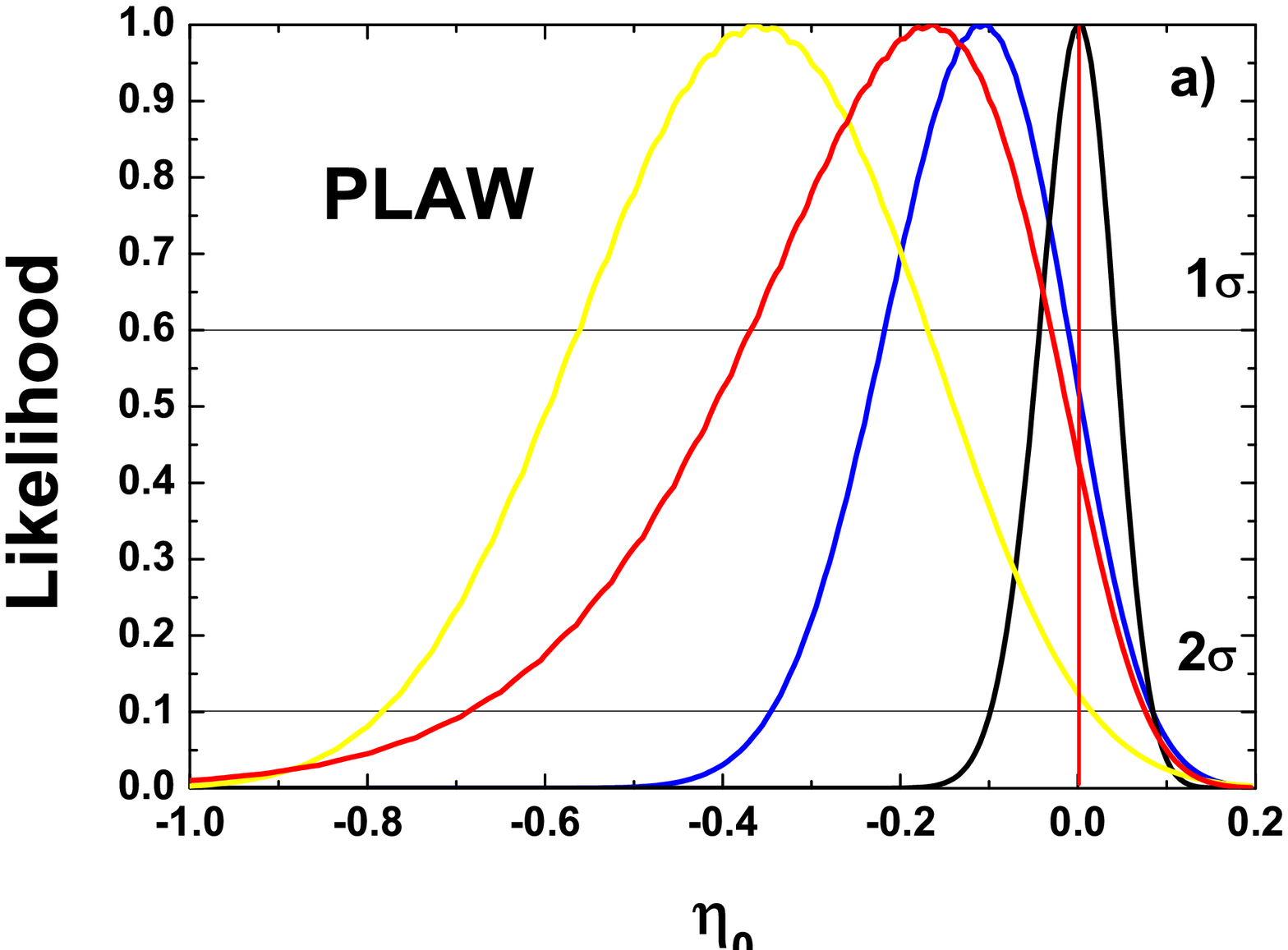}
\includegraphics[width=0.46\textwidth]{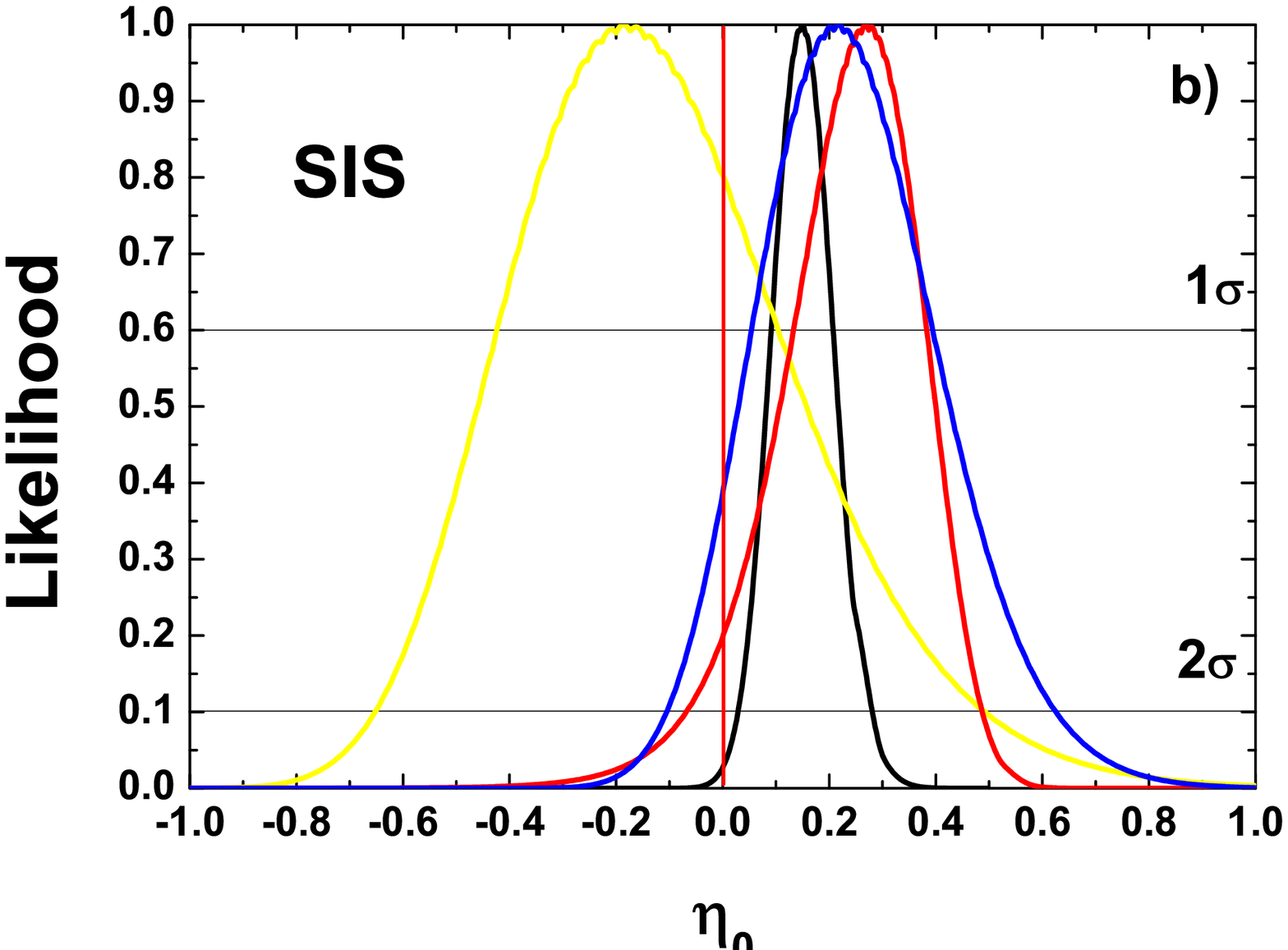}
\caption{ In fig.(a) we plot the analysis for the PLAW model. The  In fig.(b) we plot the analysis for the SIS model.{ In both figures, the solid, yellow, red and blue curves correspond to P1, P2, P3 and P4 $\eta(z)$ functions. }}
\end{figure*}

\section{Samples}

In our work, we use the following data sets:\\

\begin{itemize}
 
 \item {{\it Angular diameter distances}: we use angular diameter distances  obtained for  118 SGL systems by the Ref.~\cite{Cao}. These systems were observed in Sloan Lens ACS survey (SLACS), BOSS Emission-Line Lens Survey (BELLS), Lenses Structure and Dynamics Survey  (LSD) and  Strong Legacy Survey SL2S surveys with redshift ranges: $0.075 \leq z_l \leq 1.004$ and  $0.20 \leq z_s \leq 3.60$.} We consider the two approaches of \cite{Cao} to describe the lensing systems: {{the SIS model and another one where a spherically symmetric mass distribution in lensing galaxies is also assumed, but the rigid assumption of the SIS model is relaxed in favor of a more general power-law index $\gamma $, $\rho \propto r^{-\gamma}$ (PLAW model).}} This kind of model is important since several studies have shown that slopes of density profiles of individual galaxies show a non-negligible scatter from the SIS \cite{SIS}. Under this assumption the Einstein radius is:
\begin{equation}
 \label{Einstein} 
\theta_E =   4 \pi
\frac{\sigma_{ap}^2}{c^2} \frac{D_{ls}}{D_s} \left(
\frac{\theta_E}{\theta_{ap}} \right)^{2-\gamma} f(\gamma),
\end{equation}
where $\sigma_{ap}$ is the  stellar velocity dispersion inside the aperture of size $\theta_{ap}$ and
\begin{eqnarray} \label{f factor}
f(\gamma) &=& - \frac{1}{\sqrt{\pi}} \frac{(5-2 \gamma)(1-\gamma)}{3-\gamma} \frac{\Gamma(\gamma - 1)}{\Gamma(\gamma - 3/2)}\nonumber\\
          &\times & \left[ \frac{\Gamma(\gamma/2 - 1/2)}{\Gamma(\gamma / 2)} \right]^2.
\end{eqnarray}
Therefore: 
\begin{equation} 
\label{NewObservable}
 D=D_{A_{ls}}/D_{A_{s}} = \frac{c^2 \theta_E }{4 \pi \sigma_{ap}^2} \left( \frac{\theta_{ap}}{\theta_E} \right)^{2-\gamma} f^{-1}(\gamma).
\end{equation}
The distribution becomes a SIS for $ \gamma = 2$. In Table 1 of \cite{Cao} is displayed all  relevant information necessary to obtain $D$ in Eq.(7) for both models and perform our fit. It is important to stress that the $\sigma_{SIS}$ in Eq.(3)  is not to be exactly equal to the observed stellar velocity dispersion $\sigma_0$ since there is a strong indication that dark matter halos are dynamically hotter than the luminous stars based on X-ray observations. In order to account for this difference  it is introduced a phenomenological free parameter $f_e$, defined by the relation: $\sigma_{SIS} = f_e\sigma_0$. { In our analyses (see next section), we marginalize over $\gamma$ and $f_e$ by using the following flat priors: $(0.8)^{1/2} < f_e < (1.2)^{1/2}$ \cite{Ofe} and $\gamma$: $1.15 < \gamma < 3.5$.} Also following \cite{Cao}, we replace $\sigma_{ap}$ by $\sigma_0$ in the PLAW model. \\

\item {\it Luminosity distances}: { here we combine two data sets. The first one is the full SNe Ia sample  compiled by Suzuki et al. (2012) \cite{Suz}, the so-called Union2.1 compilation, see Fig. 1. This sample comprises 580 data points in the redshift range $0.015 \leq z \leq 1.42$. The second data set are the GRB distance moduli, $\mu(z)$,  from Demianski et al. (2016)  \cite{Dem}.} It is worth mentioning that, although details of the physical mechanism behind GRB explosions are not completely known yet, an important observational aspect of long GRBs are the several correlations between spectral and intensity properties, which suggest that GRBs can be used as distance indicators. { The authors of \cite{Dem} used SNe Ia luminosity distances (Union2.1 compilation) at redshifts close to GRBs to calibrate the Amati relation \cite{amati}, which relates the peak photon energy of a GRB $E_{p,i}$ with its isotropically equivalent radiated energy $E_{iso}$. A distance modulus to each GRB was ascribed by finding SNe Ia with small redshift distances and applying a weighted interpolation. By fitting a power law with an intrinsic scatter where SNe Ia and GRBs overlap, the parameters of the fit could be used to determine the distance modulus of the GRBs at higher redshifts and their respectively uncertainty.} A possible redshift dependence of the correlation and its effect on the GRB Hubble diagram were tested and no significant redshift dependence was found. { The original data set has 167 GRBs in the redshift range $0.033 < z < 9.3$ (see Fig. 1). We excluded 15 GRBs that are beyond the SGL systems and added to SNe Ia data the remaining 147 GRBs. Thus, we finish with 727 luminosity distances}. We obtain the luminosity distance, $D_L$, to each SNe Ia and GRB from: $D_L(z)=10^{({\mu(z)}-25)/5}$ Mpc and   $\sigma^2_{D_L}=(\frac{\partial D_L}{\partial \mu} )^2 \sigma^2_{\mu}$. { Then, we parameterize their luminosity distances  with a second degree polynomial function, such as: $D_L(z) = A z + B z^2$ (in Mpc), with $A=4488.69$, $B=1576.81$, $\sigma_A=27.06$,  $\sigma_B=77.43$ and $COV(A,B)=-1324.49216$ (see Fig. 1). Therefore, due to the high redshifts of the GRBs, the $D_L(z)$ function can provide the luminosity distances to the lens and source of each SGL system in \cite{Cao}.}

\end{itemize}

\begin{figure*}
\centering
\includegraphics[width=0.46\textwidth]{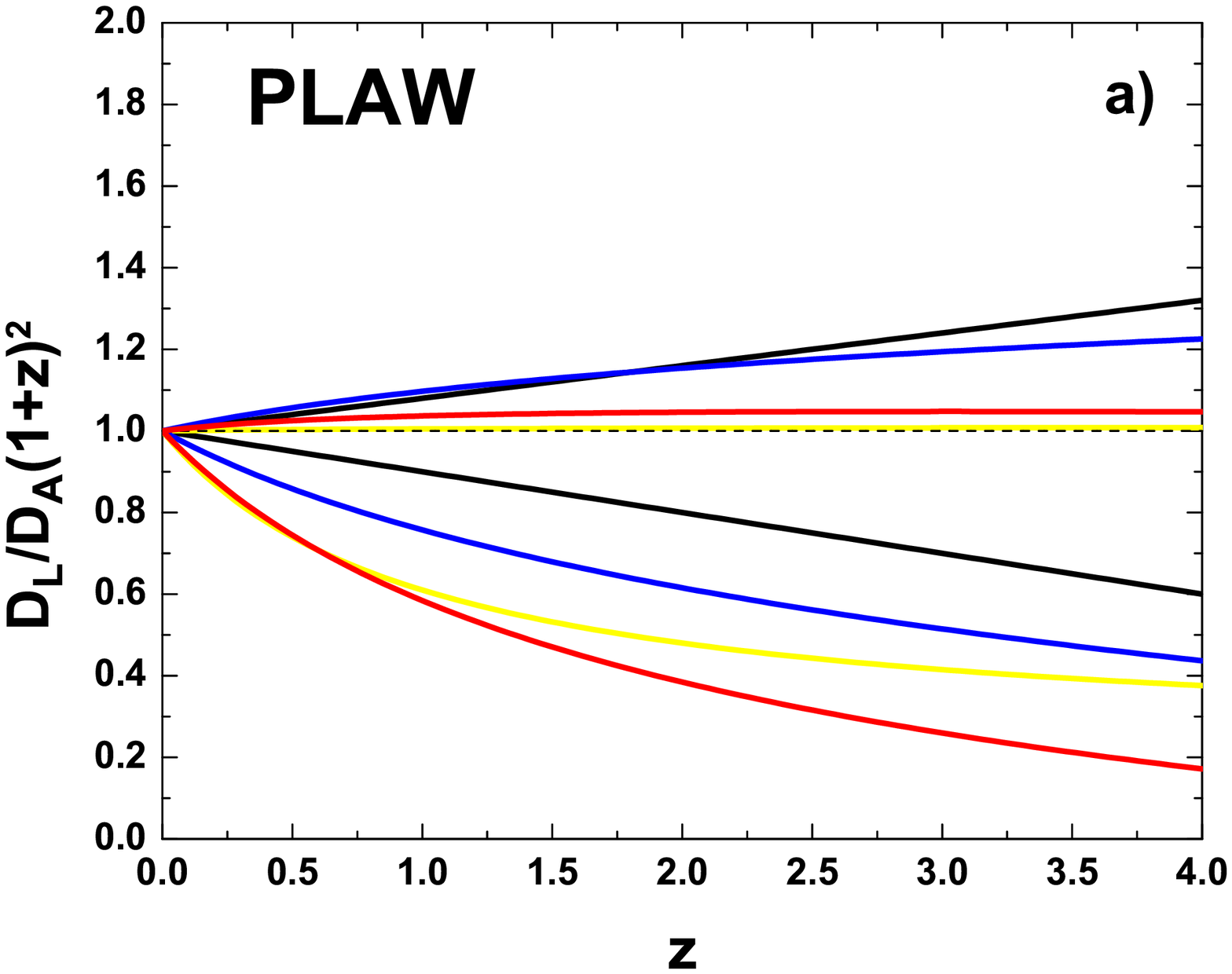}
\includegraphics[width=0.46\textwidth]{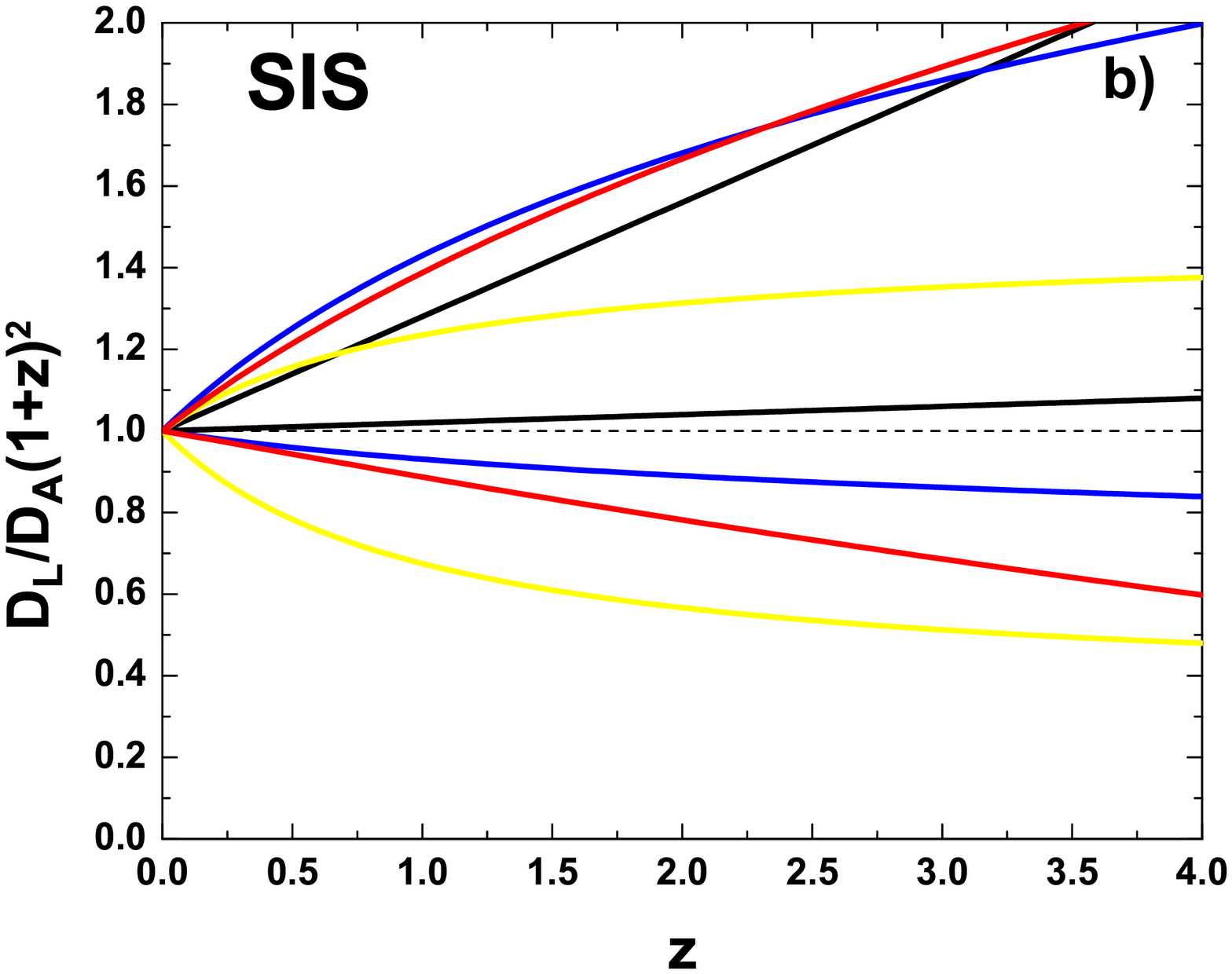}
\caption{ In fig.(a) we plot the evolution of the $\eta(z)$ functions for PLAM model. In fig.(b) we plot for SIS model.{ In both figures, the solid, yellow, red and blue curves correspond to P1, P2, P3 and P4 $\eta(z)$ functions. The curves correspond to 2$\sigma$ c.l.. }}
\end{figure*}

\begin{table*}[ht]
\caption{A summary of the current constraints on the $\eta_0$ parameters from { SGL in $2\sigma$ c.l..}}
\label{tables1}%tab2
\par
\begin{center}
\begin{tabular}{|c||c|c|c|c|c|}
\hline\hline Reference & Data Sample &$1+\eta_0z$ & $1+\eta_0z/(1+z)$&$(1+z)^{\eta_0}$&$1+\eta_0\ln (1+z)$
\\ \hline\hline 
\cite{Lia} & SGL (SIS)  + SNe Ia  & $-0.005^{+0.800}_{-0.100}$ & - & -& - \\
\cite{Hol}\footnote{Planck priors} & SGL (SIS) + SNe Ia  + $\Lambda$CDM & $0.05 \pm 0.42 $& $0.09 \pm 0.65$ & - & - \\
\cite{Hol}$^a$ & SGL (PLAW) + SNe Ia  + $\Lambda$CDM & $0.08 \pm 0.44$ & $0.06 \pm 0.67$ & - & -  \\
\cite{Hol}\footnote{WMAP9 priors} & SGL (SIS) + SNe Ia  +  $\omega(z)$CDM & $0.01 \pm 0.50 $ & $0.017 \pm 0.67$ & - & - \\
\cite{Hol}$^b$ & SGL (PLAW) + SNe Ia + $\omega(z)$CDM & $0.054 \pm 0.650$ & $0.003 \pm 0.750$& - & -\\
{This paper} & SGL (PLAW) + SNe Ia + GRBs & $0.00 \pm 0.10$ & $-0.36^{+0.37}_{-0.42}$& $-0.16^{+0.24}_{-0.51}$ & $-0.10\pm 0.24$ \\
{This paper} & SGL (SIS) + SNe Ia + GRBs & $0.15 \pm 0.13$ & $- 0.18^{+0.45}_{-0.65}$& $0.27^{+0.22}_{-0.38}$ & $0.22^{+0.40}_{-0.32}$ \\
\hline\hline
\end{tabular}
\end{center}
\end{table*}

\section{Analysis and Results } 

The constraints on the $\eta_0$ parameter are derived by evaluating the likelihood distribution 
function, ${\cal{L}} \propto e^{-\chi^{2}/2}$, with
\begin{eqnarray}
\chi^{2} & = & \sum_{i}^{118}\frac{\left[\frac{(1+z_{s_i})\eta(z_{s_i})}{(1+z_{l_i})\eta(z_{l_i})}- (1-D_i)\frac{D_{L_{s_i}}}{D_{L_{l_i}}}\right]^2}{\sigma_i^2}                                                                
\end{eqnarray}
where $\sigma_i^2$ stands for the statistical errors associated to the $D_L(z)$ function of the SNe Ia, GRBs, the gravitational lensing  observations and it is obtained by using standard propagation errors techniques. For the gravitational lensing error one may show that:
\begin{equation} \label{uncertainty}
\sigma_D = D \sqrt{4 (\delta \sigma_{0})^2 + (1-\gamma)^2 (\delta \theta_E)^2}\;,
\end{equation}
while for $D_L(z)$ function: 

\begin{eqnarray}
\sigma_{D_L}^2 & = & \left(\frac{\partial D_L}{\partial A}\right)^2\sigma^2_A +\left(\frac{\partial D_L}{\partial B}\right)^2\sigma^2_B \\ & & \nonumber + 2\left(\frac{\partial D_L}{\partial A}\frac{\partial D_L}{\partial B}\right)COV(A,B).                                                               
\end{eqnarray}
Since our $D_L(z)$ comes from a polynomial function, the points $\eta_{0_i}$ and $\eta_{0_j}$  are correlated. However, we constructed the covariance matrix and  the terms off diagonal are negligible.

{ Our results for $\eta_0$ by using the SGL models  are plotted in figures (2a) and (2b). In both figures, the solid black, yellow, red and blue curves correspond to P1, P2, P3 and P4 $\eta(z)$ functions. The black solid horizontal lines correspond to 1 and 2$\sigma$ c.l..  As one may see, differently from previous analyses in literature, our likelihoods for $\eta_0$ depend on the $\eta(z)$ functions, probably due to the high-$z$ data used in the analyses. { As one may see, the analysis using the PLAW model is  consistent with the CDRR validity only for 2$\sigma$ when P2 is used, but is verified into 1.5$\sigma$ by the other $\eta(z)$ functions.} On the other hand, using P2 the CDDR validity is verified at 1$\sigma$ for SIS model, marginally consistent for P1 and verified into 2$\sigma$ by the other $\eta(z)$ functions.  The values of $\eta_0$ are displayed  in Table I (2$\sigma$ error) along with other cosmological-model independent results using SGL.

In figures (3a) and (3b) we plot the evolution for the $\eta(z)$ functions (the curves correspond to 2$\sigma$ c.l.). The dashed line corresponds to $\eta(z)=D_LD_A^{-1}(1+z)^{-2}=1$. The curves for $\eta(z)$ functions are compatible each other, moreover, we may conclude that our results do not show significant  deviations from CDDR validity. 

 We can compare our results with methods that do not assume a cosmological model, for example (at 2$\sigma$ c.l.): Ref.~\cite{Hol2} found $\eta_0=-0.28 \pm 0.44$ and $\eta_0=-0.43 \pm 0.60$ for linear and non-linear cases by using angular diameter distances of galaxy clusters and SNe Ia. Also using angular diameter distances of galaxy clusters and SNe Ia, Ref.\cite{Li} found $\eta_0=-0.12 \pm 0.35$ and $\eta_0=-0.11 \pm 0.51$ for linear and non-linear cases. The authors of Ref.~\cite{Nair} found $\eta_0=-0.151 \pm 0.155$ at 1$\sigma$ for non-linear case by using BAO+SNe whereas Ref.~\cite{gha} found $\eta_0 = -0.08^{+2.28}_{-1.22}$ for non-linear case by using gas mass fractions of galaxy clusters and SNe Ia. The analysis performed in Ref.\cite{santos} found $\eta_0 = -0.100^{+0.117}_{-0.126}$ and $\eta_0 = -0.157^{+0.179}_{-0.192}$ for the linear and non-linear cases by using angular diameter distances  of galaxy clusters and $H(z)$ data. These authors also obtained $\eta_0 = 0.062^{+0.168}_{-0.146}$ and $\eta_0 = -0.166^{+0.337}_{-0.168}$ for the linear and non-linear cases by considering gas mass fraction measurements of galaxy clusters and $H(z)$ data. The results of Ref.~\cite{hga} point to $\eta_0=-0.15 \pm 0.25$ and $\eta_0=-0.22 \pm 0.42$ for linear and non-linear cases by using only galaxy cluster gas mass fraction measurements while Ref.~\cite{Lia} obtained $\eta_0=-0.005^{+0.800}_{-0.100}$ by using strong lensing and SNe Ia (see  table I in Ref.\cite{Hol} for recent results from several methods). The CDDR validity is verified at least at 2$\sigma$. As one may see, the method proposed here is competitive with the previous ones and includes higher-$z$ data. }

{As it is well-known,  the luminosity distance in the cosmic concordance model grows slightly faster than a quadratic function for $z > 1.5$. Thus, we also fit the SNe Ia and GRBs data by using a function that mimics such behavior and perform the analyses. We obtain that the CDDR validity ($\eta_0=0$) is verified at least within 2$\sigma$ c.l. for the SIS model and it is more than 3$\sigma$ away for the PLaw model. However, as shown in the Fig.2 where a quadratic function was considered, the PLaw model is in more agreement with the CDDR validity than the SIS model. Unlike the analyses at $z < 1$ where the results are weakly dependent on the assumed function for the luminosity distance, our results point to the fact that caution is needed when applying it to higher redshifts. Also, a better understanding of the mass distribution of lenses is required for  more accurate results.}

\section{Conclusions}

In the last few years, the main fundamental hypotheses of the standard cosmology have been tested thanks to the improved precision of cosmological data. One of them is the so-called cosmic distance duality relation involving angular diameter, $D_A$, and luminosity distances, $D_L(1+z)^{-2}D_A=\eta(z)=1$. This result is theoretically valid for all cosmological models based on the Riemannian geometry, being independent either upon Einstein field equations or the nature of matter. 

{ Nowadays,  model-independent tests using exclusively cosmological distances have been restricted up to redshifts $z = 1.50$, which it is nearly the maximum $z$ of type Ia supernovae compilations considered in analyses. In this paper, we have proposed a test that  considers 118 angular diameter distances of the strong gravitational lens systems in the redshift range $0.075 \leq z \leq 3.60$ \cite{Cao}  and { 727 luminosity distances comprising 580 type Ia supernovae from Union2.1 compilation plus 147 gamma-ray bursts with redshift lower than 3.60. These gamma-ray bursts are from a sample of 162  distance moduli with the redshift range $0.033 < z < 9.3$ from \cite{Dem}.} We have parameterized the luminosity distances of the type Ia supernovae plus gamma-ray bursts with a second degree polynomial  and used the following functions deforming the cosmic distance duality relations: (P1) $\eta(z)=1+\eta_0z$, (P2) $\eta(z)=1+\eta_0z/(1+z)$, P(3) $\eta(z)=(1+z)^{\eta_0}$ and P(4) $\eta(z)=1+\eta_0\ln(1+z)$.}

The strong gravitational lens systems were described by the simplest singular isothermal spherical model and by a more general power-law index $\gamma $, $\rho \propto r^{-\gamma}$ (PLAW model).  We have obtained that the  CDRR validity verification depends on the assumptions used to describe the lensing systems and the $\eta(z)$ functions (see Figs. 2a and 2b). For instance, when the power law model was considered, the $\eta_0$ value was compatible with $\eta_0=0$ into 1.5$\sigma$ c.l. for P1, P3 and P4  functions { and  in 2$\sigma$ for P2}. On the other hand, when the singular isothermal sphere model was used,  the CDRR validity was verified into $1.5\sigma$ level for P2, P3, P4 and only marginally consistent for P1. 

{We also fitted the SNe Ia and GRBs data by using a function that grows slightly faster than a quadratic function for $z > 1.5$. In this case, the results showed that the CDDR validity is verified at least within 2$\sigma$ c.l.  for the SIS model and it is more than 3$\sigma$ away for the PLaw model. Then, from this very first analysis at high redshifts, we conclude the results do not indicate any significant deviation from the CDDR. However, they also pointed to the fact that caution is needed when one fits data in higher redshifts and a better understanding of the mass distribution of lenses is required for more accurate results. A more in deep analysis will be presented elsewhere.}

\section*{Acknowledgments}
%----------------------------------------------------------%
RFLH is supported by INCT-A and CNPq (No. 478524/2013-7;303734/2014-0). VCB is supported by FAPESP/CAPES agreement under grant number 2014/21098-1 and FAPESP under grant number 2016/17271-5. FSL is supported by CAPES. JSA acknowledges support from CNPq and FAPERJ.  The authors are
grateful to referee for very constructive comments.


\begin{thebibliography}{99}

\bibitem{Eth} I. M. H. Etherington, Phil. Mag., {  15}, 761 (1933); reprinted in  Gen. Relativ. Gravit., {  39}, 1055 (2007)
 \bibitem{Ell} G.F.R. Ellis, R.  Poltis,  J.-P. Uzan,  A. Weltman, Phys. Rev. D, {  87}, 103530 (2013)
\bibitem{Avg} A. Avgoustidis, C. Burrage, J. Redondo, L. Verde and R. Jimenez, JCAP, {  10}, 024 (2010); A. Avgoustidis, G. Luzzi,  C. J. A. P. Martins and  A. M. R. V. L. Monteiro, JCAP, {  2}, 013 (2012)
 \bibitem{Dis1}B. A. Bassett and M. Kunz, 
Phys. Rev. D {  69}, 101305 (2004); J. P. Uzan, N. Aghanim \& Y. Mellier, Phys. Rev. D, {  70}, 083533 (2004); R. F. L. Holanda, J. A. S. Lima and M. B. Ribeiro, A\&A, {  528}, L14 (2011)
R. F. L. Holanda and V. C. Busti, Phys. Rev. D {  89}, 103517 (2014)  [arXiv:1402.2161] ;   
\bibitem{dis2} Xiao-Lei Meng, Tong-Jie Zhang , Hu Zhan and Xin Wang, APJ, {  745}, 98 (2012);
 X. Yang et. al., 
%\emph{An improved method to test the Distance--Duality relation}, 
Astrophys. J. Lett. {  777}, L24 (2013); G.F.R. Ellis, R. Poltis,  J.-P. Uzan and A. Weltman, Phys. Rev. D, {  87}, 103530 (2013); A. Rana, D. Jain, S. Mahajan and A. Mukherjee, JCAP, {  07}, 026 (2016)
\bibitem{hga}R. F. L. Holanda, R. S. Gon\c{c}alves and J. S. Alcaniz,  
JCAP 1206 , 022 (2012) [arXiv:1201.2378]
\bibitem{Li}Z. Li, P. Wu and W. Yu, 
Astrophys. J. {  729}, L14 (2011);
\bibitem{gha}R. S. Gon\c{c}alves, R. F. L. Holanda and J. S. Alcaniz, 
Mon. Not. Roy. Astron. Soc. {  420} , L43 (2012) [arXiv:1109.2790]
\bibitem{santos}S. Santos-da-Costa, V. C. Busti and R. F. L. Holanda, JCAP, {  10}, 061 (2015)
\bibitem{Hol2}R. F. L. Holanda, J. A. S. Lima and M. B. Ribeiro, ApJL, {  722}, 233 (2010)
\bibitem{Nair} R. J. Nair, S. and  D. Jain, JCAP, {  05}, 023 (2005)
\bibitem{Hol}R. F. L. Holanda,  V. C. Busti and  J. S. Alcaniz, JCAP, {  02}, 054 (2016) 
\bibitem{Lia}W. Liao et al., ApJL, {  822}, 74 (2016) 
\bibitem{SGL}Z.H. Zhu, Mod. Phys. Lett. A, {  15}, 1023 (2000);  K.-H. Chae and S. D. Mao, ApJ, {  599}, L61 (2003); J. L. Mitchell, C. R. Keeton,  J. A. Frieman and  R. K. Sheth, ApJ, {  622}, 81 (2005);  Z.-H. Zhu and S. Mauro, A\&A, {  487}, 831 (2008); Z.-H. Zhu et al., A\&A, {  483}, 15 (2008); K.-H. Chae,  G. Chen,  B. Ratra and D.-W. Lee, ApJ, {  607}, L71 (2004); M. Biesiada, B. Malec and A. Piorkowska, MNRAS, {  406},1055 (2010);  C. C. Yuan and  F. Y. Wang, MNRAS, {  452}, 2423 (2015); E. V. Linder, Phys. Rev. D, {  94}, 083510 (2016)
\bibitem{Sch}Schneider, P., Kochanek, C. S., \& Wambsganss, J. 2006, Gravitational Lensing: Strong, Weak and Micro (Springer)
\bibitem{Cao} S. Cao, M. Biesiada, R. Gavazzi, A. Piorkowska and Z. U. Zhu, ApJ, {  806}, 185 (2015) 
\bibitem{Suz}N. Suzuki et al., ApJ, {  85}, 746 (2012)
\bibitem{Ade}P. A. R. Ade et al., A\&A, {  594}, A13 (2016)
\bibitem{Hin}G. Hinshaw et al., APJS, {  208}, 19H (2013)
\bibitem{Bet}M. Betoule et. al., A\&A, {  568}, A22 (2014)
\bibitem{Dem} M. Demianski, E. Piedipalumbo, D. Sawant and  L. Amati, A\&A, {  598}, A112 (2017)
\bibitem{Bal}M. Bartelmann and P. Schneider, Phys. Rep., {  340}, 291 (2001)
\bibitem{SIS}L. Koopmans, A. Bolton, T. Treu, O. Czoske, M. Auger, et al., ApJ, {  703}, L54 (2009); M. W. Auger, T. Treu, A. S. Bolton, R. Gavazzi, L. V. E. Koopmans, P. J. Marshall, L. A. Moustakas and S. Burles, ApJ, {  724} 511 (2010); M. Barnabe, O. Czoske, L. V. E. Koopmans, T. Treu and A. S. Bolton, MNRAS, {  415}, 2215 (2011); A. Sonnenfeld, T. Treu, R. Gavazzi, S. H. Suyu, P. J. Marshall, et al., ApJ, {  777}, 98 (2013); M. Cappellari et al. 2015 arXiv:1504.0007
\bibitem{Ofe} E. O. Ofek, H. W. Rix and D. Maoz, MNRAS, {  343}, 639 (2003)
\bibitem{amati} L. Amati {\it et al.}, A\&A, {  390}, 81 (2002)









\end{thebibliography}
\end{document}